\documentclass[12pt]{article}
\usepackage{makeidx}
\usepackage{amssymb}
\usepackage{amsfonts}
\usepackage{amsmath}
\usepackage{geometry}
\usepackage[colorlinks=true, linkcolor=red, citecolor=blue]{hyperref}
\usepackage{float}
\usepackage{mathtools}
\usepackage{caption}
\usepackage{graphicx}
\usepackage[font=small,labelfont=bf,labelsep=space]{caption}
\usepackage{subfig}
\usepackage{setspace}
\usepackage{physics}
\usepackage{commath}
\usepackage{xcolor}
\usepackage{algorithm}
\usepackage[noend]{algpseudocode}

\makeatletter

\@addtoreset{equation}{section}
\makeatother 

\newtheorem{theorem}{Theorem}

\newtheorem{definition}[theorem]{Definition}

\newtheorem{lemma}[theorem]{Lemma}

\newcommand{\beq}{\begin{eqnarray}}
\newcommand{\eeq}{\end{eqnarray}}

\begin{document}

\title{Resurgence Theory and Holomorphic Quantum Mechanics}
\author{M. W. AlMasri\\
{\footnotesize Wilczek Quantum Center, School of Physics and Astronomy,}\\
{\footnotesize Shanghai Jiao Tong University, Minhang, Shanghai, China}\\
{\footnotesize e-mail: mwalmasri2003@gmail.com}}
\maketitle

\begin{abstract}
In this work, we study the resurgence program in holomorphic quantum mechanics. As a specific problem, we investigate the resurgence in the quartic anharmonic oscillator within holomorphic quantum mechanics, using the Bargmann representation of bosonic operators. In this framework, the perturbative energy series is shown to be Gevrey-1 and Borel summable only after continuation across the Stokes line. The instanton operator, realized as a coherent-state displacement in the Segal--Bargmann space, provides an explicit operatorial bridge between perturbative coefficients and non-perturbative sectors. Alien derivative relations generate the full resurgence triangle characteristic of the Bender–Wu model, and the resummed energy is expressed as a trans-series via a ratio of expectation values involving this instanton operator. As a concrete demonstration, we compute the first seven energy levels (\(n=0,\dots,6\)) up to sixth order in the coupling \(g\); the resulting exact rational coefficients reproduce the classic Bender–Wu results, confirming the consistency and power of the holomorphic resurgence approach.
\end{abstract}

\section{Introduction}
Quantum field theory (QFT) stands as the foundational framework for our understanding of fundamental interactions in particle physics, statistical mechanics, and condensed matter systems \cite{Peskin,Zinnjustin,Fradkin}. Despite its success, a complete nonperturbative formulation of most interacting QFTs remains elusive. Perturbative methods—based on asymptotic expansions in a small coupling constant—constitute the primary computational tool in practice; however, such expansions are generically divergent. This divergence reflects the intrinsic limitations of perturbation theory and signals the necessity of incorporating nonperturbative effects—such as instantons, renormalons, and other semiclassical configurations—for a consistent and complete description of the theory. Over the past two decades, resurgence theory has emerged as a rigorous and unifying framework capable of bridging perturbative and nonperturbative sectors of quantum theories \cite{Marino}. Originating in the mathematical work of \'Ecalle and later by Berry and Howls, resurgence provides a systematic method for analyzing the analytic structure of asymptotic series \cite{Ecalle,Berry}. In the context of quantum field theory, this program has been decisively advanced by the contributions of Dunne, \"Unsal, and collaborators, who have demonstrated that the large-order behavior of perturbative coefficients encodes precise information about nonperturbative saddle points of the path integral \cite{Dorigoni,Dunne}.

Central to this approach is the concept of a \emph{trans-series}, an extended asymptotic expansion that incorporates not only powers of the coupling but also nonanalytic exponential terms (e.g., \(e^{-S/g}\), where \(S\) is the action of a semiclassical saddle). Resurgence theory posits that the ambiguities inherent in the Borel resummation of perturbative series are exactly compensated by ambiguities in the nonperturbative sectors, yielding unambiguous physical observables. This cancellation is governed by algebraic and analytic relations—known as \emph{resurgence relations}—that tightly constrain the global structure of the theory. Later works have firmly established resurgence as a concrete and computationally viable framework in quantum field theory through detailed analyses of quantum mechanical models, two-dimensional sigma models, four-dimensional gauge theories, and many others \cite{Unsal,Dunne0,Dunne1,Tin}.

One interesting reformulation of quantum mechanics based on entire functions was suggested by Segal and Bargmann in \cite{Segal,Bargmann}. In the Bargmann (or Bargmann--Fock) representation, the creation operator $a^\dagger$ acts as multiplication by $z$, and the annihilation operator $a$ acts as differentiation $\frac{d}{dz}$ i.e.
\[
(a f)(z) = \frac{d}{dz} f(z), \quad (a^\dagger f)(z) = z f(z)
\]
\cite{Segal,Bargmann,Perelomov,Folland,Hall}. The complex variable $z$ can be interpreted as a dimensionless combination of position and momentum:
\[
z = \frac{1}{\sqrt{2}} \left( \frac{x}{\ell} + i \frac{\ell p}{\hbar} \right), \quad \text{where } \ell = \sqrt{\frac{\hbar}{m\omega}}.
\]
This identifies $z$ with a point in classical phase space, and the entire formalism becomes a form of phase space quantization. Therefore, Bargmann representation provides a rigorous setting for coherent state quantization \cite{Perelomov}. The Bargmann representation has been applied to a wide range of physical problems, including the WKB method \cite{Voros}, thermal coherent states \cite{Bishop}, the Jaynes-Cummings-Gaudin model \cite{Talatev}, quantum absorption refrigerators \cite{almasri}, damped spin systems \cite{master}, quantum computation \cite{Chabaud}, PT-symmetric Hamiltonians \cite{Wahiddin}, neural networks \cite{neuron}, logic gates \cite{complex}, the central spin model coupled to a classical Ising bath \cite{PLA}, and many other studies related to coherent states and quantization in phase space. In this work, our aim is to study the resurgence in holomorphic quantum theory of Segal and Bargmann.

The paper is organized as follows. In Section~\ref{sec:formalism}, we review the formalism of holomorphic quantum mechanics and outline the key elements of the resurgence program. Section~\ref{sec:results} presents the main results of this work: an exact perturbative and resurgent analysis of the quartic anharmonic oscillator in the Segal–Bargmann space. We conclude with a summary and outlook in Section~\ref{sec:conclusion}.

\section{Formalism}\label{sec:formalism}
\subsection{Holomorphic Quantum Mechanics}
\begin{definition}[Segal--Bargmann space]\label{def:SB}
The \emph{Segal--Bargmann space}, denoted $\mathcal{H}_{\mathrm{SB}}$, is the Hilbert space of entire functions on $\mathbb{C}$ that are square-integrable with respect to the Gaussian weight $e^{-|z|^{2}}$. Explicitly,
\begin{equation}
\mathcal{H}_{\mathrm{SB}}
:=
\left\{
f : \mathbb{C} \to \mathbb{C}
\;\middle|\;
f \text{ is entire and }
\|f\|^{2}
:=
\int_{\mathbb{C}} |f(z)|^{2}\, e^{-|z|^{2}}\, d^{2}z
< \infty
\right\},
\end{equation}
equipped with the inner product
\begin{equation}
\langle f, g \rangle
:=
\int_{\mathbb{C}} \overline{f(z)}\, g(z)\, e^{-|z|^{2}}\, d^{2}z.
\end{equation}
\end{definition}
An orthonormal basis of $\mathcal{H}_{\mathrm{SB}}$ is given by the normalized monomials
\begin{equation}\label{basis}
\phi_{n}(z) = \frac{z^{n}}{\sqrt{n!}}, \qquad n = 0,1,2,\dots .
\end{equation}

\begin{lemma}[Unitary equivalence via the Segal--Bargmann transform]\label{lem:SB_unitary}
The Segal--Bargmann space $\mathcal{H}_{\mathrm{SB}}$ is unitarily isomorphic to the Schr\"odinger Hilbert space $L^{2}(\mathbb{R})$. The unitary isomorphism is given by the Segal--Bargmann transform
\begin{equation}
(T\psi)(z)
=
\frac{1}{\pi^{1/4}}
\int_{\mathbb{R}}
\exp\!\left(
-\frac{1}{2}z^{2}
-\frac{1}{2}x^{2}
+ \sqrt{2}\, x z
\right)\,
\psi(x)\, dx,
\qquad \psi \in L^{2}(\mathbb{R}),
\label{eq:SB_transform}
\end{equation}
which satisfies
\begin{equation}
\|T\psi\|_{\mathcal{H}_{\mathrm{SB}}} = \|\psi\|_{L^{2}(\mathbb{R})}
\quad \text{for all } \psi \in L^{2}(\mathbb{R}),
\end{equation}
and extends uniquely to a unitary operator $T : L^{2}(\mathbb{R}) \to \mathcal{H}_{\mathrm{SB}}$.
\end{lemma}

In Segal-Bargmann space $\mathcal{H}_\text{SB}$, the annihilation operator acts as:
\begin{equation}
a f(z) = \frac{d}{dz} f(z)
\end{equation}
and the creation operator acts as:
\begin{equation}
a^\dagger f(z) = z f(z)
\end{equation}
They satisfy the canonical commutation relation:
\begin{equation}
[a, a^\dagger] = 1
\end{equation}
In standard quantum mechanics, the Hamiltonian of the harmonic oscillator is:
\begin{equation}
H = \frac{1}{2}(p^2 + x^2)
\end{equation}
In terms of creation and annihilation operators (with $\hbar = m = \omega = 1$):
\begin{equation}
H = a^\dagger a + \frac{1}{2}
\end{equation}
So in Segal--Bargmann space, the Hamiltonian becomes:
\begin{equation}
H = z \frac{d}{dz} + \frac{1}{2}
\end{equation}
This is a differential operator acting on analytic functions $f(z)$ in the Segal--Bargmann space.

The eigenfunctions of the quantum harmonic oscillator in position space are given in terms of Hermite polynomials. The $ n^\text{th} $-energy eigenstate is:
\begin{equation}
\psi_n(x) = \left( \frac{1}{\sqrt{\pi} 2^n n!} \right)^{1/2} e^{-x^2/2} H_n(x)
\end{equation}
where $ H_n(x) $ is the $ n^\text{th} $-degree Hermite polynomial.
These functions form an orthonormal basis in $ L^2(\mathbb{R}) $, and satisfy the time-independent Schrödinger equation:
\begin{equation}
-\frac{1}{2} \frac{d^2}{dx^2} \psi_n(x) + \frac{1}{2} x^2 \psi_n(x) = \left(n + \frac{1}{2}\right) \psi_n(x)
\end{equation}
The Hermite polynomials $ H_n(x) $ are defined via the Rodrigues' formula:
\begin{equation}
H_n(x) = (-1)^n e^{x^2} \frac{d^n}{dx^n} \left( e^{-x^2} \right)
\end{equation}
The Segal--Bargmann transform maps the $ n^\text{th} $ Hermite function to the monomial basis of Segal--Bargmann space:
\begin{equation}
T\psi_n(z) = \phi_n(z) = \frac{z^n}{\sqrt{n!}}
\end{equation}
The inverse transform recovers the original wavefunction is
\begin{equation}
(T^{-1}f)(x) = \int_{\mathbb{C}} e^{-\frac{x^2}{2} - \frac{z^2}{2} + x \sqrt{2} z} f(z) \, e^{-|z|^2} d^2z
\end{equation}
The time-dependent Schrödinger equation is:
\begin{equation}
i \frac{\partial}{\partial t} \psi(z,t) = H \psi(z,t)
\end{equation}
Substituting the expression for $H$, we get:
\begin{equation}
i \frac{\partial}{\partial t} \psi(z,t) = \left( z \frac{\partial}{\partial z} + \frac{1}{2} \right) \psi(z,t)
\end{equation}
This is a first-order linear partial differential equation in time and complex variable $z$.
We can solve this by expanding $\psi(z,t)$ in the orthonormal basis $\phi_n(z) $, since these are eigenfunctions of $H$:
\[
H \phi_n(z) = \left(n + \frac{1}{2}\right) \phi_n(z)
\]
So if the initial wavefunction is:
\[
\psi(z,0) = \sum_{n=0}^\infty c_n \phi_n(z)
\]
Then the time evolution is:
\begin{equation}
\psi(z,t) = \sum_{n=0}^\infty c_n e^{-i(n + \frac{1}{2})t} \phi_n(z)
\end{equation}
This gives the full solution to the Schrödinger equation in Segal--Bargmann space.

\begin{lemma}[Reproducing kernel of the Segal--Bargmann space]\label{lem:reproducing_kernel}
The Segal--Bargmann space $\mathcal{H}_{\mathrm{SB}}$ is a reproducing kernel Hilbert space. Its reproducing kernel is the entire function
\begin{equation}
K(z,w) = e^{\,z\bar{w}}, \qquad z,w \in \mathbb{C},
\end{equation}
which admits the series representation
\begin{equation}
K(z,w) = \sum_{n=0}^{\infty} \phi_n(z)\,\overline{\phi_n(w)}
= \sum_{n=0}^{\infty} \frac{z^n \bar{w}^n}{n!},
\end{equation}
where $\{\phi_n\}_{n=0}^\infty$, $\phi_n(z) = z^n/\sqrt{n!}$, is the orthonormal monomial basis of $\mathcal{H}_{\mathrm{SB}}$.
Moreover, $K$ satisfies the reproducing property
\begin{equation}
\langle K(\cdot, w),\, \psi \rangle_{\mathcal{H}_{\mathrm{SB}}} = \psi(w),
\qquad \forall\, w \in \mathbb{C},\; \forall\, \psi \in \mathcal{H}_{\mathrm{SB}}.
\end{equation}
\end{lemma}

The Husimi Q-function is a quasiprobability distribution used to represent quantum states in phase space. Thus, computing the Husimi $Q$-function in Segal--Bargmann space simplifies the calculations since we deal with position and momentum in one complex variable. For a normalized wavefunction $\psi(z)$ in Segal--Bargmann space, the Husimi function is defined as:
\begin{equation}
Q_\psi(z) = \frac{1}{\pi} e^{-|z|^2} |\psi(z)|^2
\end{equation}
This gives a non-negative probability density over the complex plane. It's essentially the modulus squared of the wavefunction weighted by the Gaussian factor $e^{-|z|^2}$.

\begin{definition}[Toeplitz operator]\label{def:Toeplitz}
Let $H$ be a Hilbert space of (complex-valued) functions on a measure space, and let $K \subseteq H$ be a closed subspace with orthogonal projection $P \colon H \to K$. For a bounded measurable function $\phi \colon X \to \mathbb{C}$ (called the \emph{symbol}), the \emph{Toeplitz operator} with symbol $\phi$ is the linear operator $T_\phi \colon K \to K$ defined by
\begin{equation}
T_\phi f = P(\phi f), \qquad f \in K.
\end{equation}
In words, $T_\phi$ acts by pointwise multiplication by $\phi$ in the ambient space $H$, followed by orthogonal projection onto the subspace $K$ \cite{Douglas}.
\end{definition}

In Segal--Bargmann space $ \mathcal{H}_\text{SB} $, a Toeplitz operator is defined as follows: given a measurable function $ f(z)$ (the symbol), the associated Toeplitz operator $ T_f $ acts on any $ \psi(z) \in \mathcal{H}_\text{SB} $ as:
\begin{equation}
(T_f \psi)(z) = P_\text{SB} \left( f(z) \psi(z) \right)
\end{equation}
where:
$ P_\text{SB}: L^2(\mathbb{C}, e^{-|z|^2}) \to \mathcal{H}_\text{SB} $ is the orthogonal projection onto the Segal--Bargmann space,
$ f(z) $ is the function representing an observable. This means we multiply $ \psi(z) $ by $ f(z) $, then project back into the Segal--Bargmann space.
Let's define the matrix elements of $ T_f $ in this basis:
\begin{equation}\label{tmatrix}
(T_f)_{mn} = \langle \phi_m | T_f | \phi_n \rangle
= \int_{\mathbb{C}} \overline{\phi_m(z)} f(z) \phi_n(z) e^{-|z|^2} d^2z
\end{equation}
Substituting the basis functions defined in Eq.\ref{basis}, we obtain:
\begin{equation}
(T_f)_{mn} = \frac{1}{\sqrt{m!n!}} \int_{\mathbb{C}} \bar{z}^m f(z) z^n e^{-|z|^2} d^2z
\end{equation}
This gives us a practical way to compute Toeplitz operators numerically or analytically.
Let's take $ f(z) = z $. Then the associated Toeplitz operator acts as:
\begin{equation}
T_z \psi(z) = P_\text{SB} \left( z \psi(z) \right)
\end{equation}
where $ P_\text{SB} $ is the orthogonal projection onto Segal--Bargmann space.
If $ \psi(z) $ is already analytic (i.e., in Segal--Bargmann space), then multiplying by $ z $ keeps it analytic. So in this case, projection does nothing — the result stays in Segal--Bargmann space.
Therefore,
\begin{equation}
T_z \psi(z) = z \psi(z)
\end{equation}
This means the operator $ T_z $ is just multiplication by $ z $ and it's a holomorphic Toeplitz operator. It represents the position operator in complex representation and is one of the simplest examples of a Toeplitz operator.

\begin{theorem}[Ladder operators as Toeplitz operators \cite{Hall}]\label{thm:ladder_Toeplitz}
Let $\mathcal{H}_{\mathrm{SB}} \subset L^2(\mathbb{C}, e^{-|z|^2}\,d^2z)$ be the Segal--Bargmann space, and let $P \colon L^2(\mathbb{C}, e^{-|z|^2} d^2z) \to \mathcal{H}_{\mathrm{SB}}$ denote the orthogonal Bargmann projection. For $f \in L^\infty(\mathbb{C})$, define the Toeplitz operator $T_f$ on $\mathcal{H}_{\mathrm{SB}}$ by $T_f \psi = P(f\psi)$. Then the creation and annihilation operators are realized as Toeplitz operators with linear symbols:
\begin{align}
a^\dagger &= T_z, \\
a        &= T_{\bar{z}}.
\end{align}
That is, for all $\psi \in \mathcal{H}_{\mathrm{SB}}$,
\[
a^\dagger \psi(z) = P\!\bigl(z\, \psi(z)\bigr) = z\,\psi(z), \qquad
a\, \psi(z) = P\!\bigl(\bar{z}\, \psi(z)\bigr) = \partial_z \psi(z) .
\]
\end{theorem}
While $ T_z $ acts directly as multiplication (no projection needed), $ T_{\bar{z}} $ requires projection back into the Segal--Bargmann space since $ \bar{z} $ is not analytic.

Let's consider $ f(z) = |z|^2 $. This corresponds to the classical Hamiltonian of the harmonic oscillator. Then the matrix elements of $T_{f}$ is:
\begin{equation}
(T_f)_{mn} = \frac{1}{\sqrt{m!n!}} \int_{\mathbb{C}} \bar{z}^m |z|^2 z^n e^{-|z|^2} d^2z
= \frac{1}{\sqrt{m!n!}} \int_0^{2\pi} \int_0^\infty r^{m+n+2} e^{i(n - m)\theta} e^{-r^2} r dr d\theta
\end{equation}
The angular integral gives zero unless $ m = n $, so only diagonal entries survive:
$$(T_f)_{nn} = \frac{1}{n!} \int_0^\infty r^{2n + 2} e^{-r^2} r dr \cdot \int_0^{2\pi} d\theta
= \frac{2\pi}{n!} \cdot \frac{1}{2} \Gamma(n + 2)
= \frac{\pi}{n!} \cdot (n + 1)! = \pi (n + 1)$$
Thus:
\begin{equation}
(T_f)_{mn} = \pi (n + 1) \delta_{mn}
\end{equation}
This is proportional to the quantum harmonic oscillator Hamiltonian $H \phi_n(z) = \left(n + \frac{1}{2}\right) \phi_n(z)$.
So up to constants and normalization, the harmonic oscillator Hamiltonian is a Toeplitz operator with function $ f(z) = |z|^2 $.

A coherent state in Segal--Bargmann space is:
$$
\psi_\alpha(z) = e^{\alpha z - \frac{|\alpha|^2}{2}} = e^{-\frac{|\alpha|^2}{2}} e^{\alpha z}
$$
It satisfies:
\begin{equation}
a \psi_\alpha(z) = \alpha \psi_\alpha(z)
\end{equation}
Its time evolution under the Hamiltonian is:
\begin{equation} \psi_\alpha(z,t) = e^{-iHt} \psi_\alpha(z) = e^{-i\frac{t}{2}} \psi_{\alpha e^{-it}}(z)= e^{-\frac{|\alpha|^2}{2}} e^{-i\frac{t}{2}} e^{\alpha e^{-it} z}
\end{equation}
This shows that the coherent state remains a coherent state during evolution, with its center rotating in the complex plane $
\alpha(t) = \alpha e^{-it}.
$
Then Husimi $Q$-function for coherent state is:
\begin{eqnarray}
Q_\alpha(z) = \frac{1}{\pi} e^{-|z|^2} |\psi_\alpha(z)|^2 = \frac{1}{\pi} e^{-|z|^2} \left| e^{-\frac{|\alpha|^2}{2}} e^{\alpha z} \right|^2\nonumber\\
= \frac{1}{\pi} e^{-|z|^2} e^{-|\alpha|^2} e^{\alpha z + \bar{\alpha} \bar{z}} = \frac{1}{\pi} e^{-|z|^2} e^{-|\alpha|^2} e^{2\,\text{Re}(\alpha z)} = \frac{1}{\pi} e^{-|z - \alpha|^2}
\end{eqnarray}
This is a Gaussian centered at $z = \alpha$ just like phase space.
Finally, we compute the Toeplitz operator for the coherent state $\psi_{\alpha}(z)$. We plug the coherent state into the formula Eq. \ref{tmatrix}. We obtain:
\begin{equation}
(T_{\psi_\alpha})_{mn} = \frac{e^{-\frac{|\alpha|^2}{2}}}{\sqrt{m!n!}} \int_{\mathbb{C}} \bar{z}^m z^n e^{\alpha z} e^{-|z|^2} d^2z
\end{equation}
In order to evaluate the previous integral, we switch to polar coordinates:
\begin{equation}
z = r e^{i\theta}, \quad d^2z = r dr d\theta
\end{equation}
Then:
\begin{equation}
\bar{z}^m z^n e^{\alpha z} = r^{m+n} e^{-im\theta + in\theta + \alpha r e^{i\theta}} = r^{m+n} e^{i(n - m)\theta} e^{\alpha r e^{i\theta}}
\end{equation}
The full expression becomes:
$$
(T_{\psi_\alpha})_{mn} = \frac{e^{-\frac{|\alpha|^2}{2}}}{\sqrt{m!n!}} \int_0^\infty \int_0^{2\pi} r^{m+n} e^{i(n - m)\theta} e^{\alpha r e^{i\theta}} e^{-r^2} r dr d\theta
$$
Now we expand $ e^{\alpha r e^{i\theta}} $ using its Taylor series
$
e^{\alpha r e^{i\theta}} = \sum_{k=0}^\infty \frac{(\alpha r)^k e^{ik\theta}}{k!}
$
and substitute into the integral:
\begin{equation}
(T_{\psi_\alpha})_{mn} = \frac{e^{-\frac{|\alpha|^2}{2}}}{\sqrt{m!n!}} \sum_{k=0}^\infty \frac{\alpha^k}{k!} \int_0^\infty r^{m+n+k+1} e^{-r^2} dr \int_0^{2\pi} e^{i(n - m + k)\theta} d\theta
\end{equation}
The angular integral gives zero unless $ n - m + k = 0 $, i.e., $ k = m - n $. Thus, only one term survives in the sum:
\begin{equation}
(T_{\psi_\alpha})_{mn} = \frac{e^{-\frac{|\alpha|^2}{2}}}{\sqrt{m!n!}} \cdot \frac{\alpha^{m - n}}{(m - n)!} \cdot \int_0^\infty r^{2m + 1} e^{-r^2} dr \cdot 2\pi
\end{equation}
Using
$
\int_0^\infty r^{2m + 1} e^{-r^2} dr = \frac{1}{2} \Gamma(m + 1) = \frac{m!}{2}
$, the radial integral can be simplified further,
\begin{equation}
(T_{\psi_\alpha})_{mn} =
\begin{cases}
\displaystyle \frac{e^{-\frac{|\alpha|^2}{2}}}{\sqrt{m!n!}} \cdot \frac{\alpha^{m - n}}{(m - n)!} \cdot m! \pi & \text{if } m \geq n \\
0 & \text{if } m < n
\end{cases}
\end{equation}
Or more compactly:
\begin{equation}
(T_{\psi_\alpha})_{mn} = \pi \cdot \delta_{m \geq n} \cdot \frac{e^{-\frac{|\alpha|^2}{2}}}{\sqrt{m!n!}} \cdot \frac{\alpha^{m - n}}{(m - n)!} \cdot m!
\end{equation}
The matrix element is non-zero only when $ m \geq n $.
It contains a factor of $ \alpha^{m - n} $ indicating overlap between Fock states induced by the coherent state and the exponential prefactor ensures normalization.

\subsection{Resurgence Program}
Perturbation theory in quantum field theory and quantum mechanics typically yields formal power series that are \emph{asymptotic} rather than convergent. For a coupling constant $g$, the perturbative expansion of an observable $\mathcal{O}(g)$ takes the form
\begin{equation}
\mathcal{O}(g) \;\sim\; \sum_{k=0}^{\infty} \mathcal{O}^{(k)}\,g^{k},
\qquad g \to 0^{+},
\end{equation}
where the coefficients $\mathcal{O}^{(k)}$ grow factorially,
\begin{equation}\label{oseries}
\mathcal{O}^{(k)} \;\sim\; (-1)^{k}\,A\,k!\,k^{b}\quad (k\to\infty).
\end{equation}
The constant $A>0$ is the \emph{instanton action} (the exponential weight of the leading non‑perturbative saddle), while $b$ is a subleading exponent encoding fluctuations around the saddle.
Because of the factorial growth, the series \ref{oseries} is not Borel summable on the positive real axis: its Borel transform
\begin{equation}
\widehat{\mathcal{O}}(\xi) \;=\; \sum_{k=0}^{\infty} \frac{\mathcal{O}^{(k)}}{k!}\,\xi^{k}
\end{equation}
possesses singularities on the positive real $\xi$–axis at $\xi = \pm A$. These singularities encode the very same non‑perturbative physics that the original perturbative series misses.

Resurgence is the framework that systematically relates the large‑order behaviour of perturbative coefficients to the data of non‑perturbative sectors (instantons, renormalons, etc.). The central objects are:
\begin{itemize}
\item The \emph{Borel transform} $\widehat{\mathcal{O}}(\xi)$,
\item The \emph{alien derivative} $\Delta_{\omega}$, which extracts the discontinuity of $\widehat{\mathcal{O}}$ across a singularity at $\xi=\omega$,
\item The \emph{resurgent trans‑series}, which combines all sectors into a single analytic object.
\end{itemize}
For a system with a single instanton action $S_{\text{inst}}$, the resurgent trans‑series for an energy level $E_{n}(g)$ reads
\begin{equation}
E_{n}(g;\sigma) \;=\;
\sum_{\ell=0}^{\infty} \sigma^{\ell}\,
e^{-\ell S_{\text{inst}}/g}\,
g^{\ell b_{n}}\,
\Phi_{n}^{(\ell)}(g),
\qquad
\Phi_{n}^{(0)}(g)=\sum_{k=0}^{\infty}E_{n}^{(k)}g^{k}.
\end{equation}
Here $\sigma$ is a Stokes parameter which is fixed by boundary conditions, and $\Phi_{n}^{(\ell)}(g)$ are perturbative expansions around the $\ell$‑instanton background.
The alien derivative connects adjacent sectors:
\begin{equation}
\Delta_{S_{\text{inst}}}\,\Phi_{n}^{(\ell)} \;=\; (\ell+1)\,C_{n}\,\Phi_{n}^{(\ell+1)},
\end{equation}
where $C_{n}$ is a Stokes constant that can be computed from exact WKB or from the large‑order asymptotics of $E_{n}^{(k)}$.

\section{Results}\label{sec:results}
We consider the quartic anharmonic oscillator in units $\hbar = m = \omega = 1$,
\begin{equation}
H(g) = \frac{1}{2}\bigl(p^{2} + x^{2}\bigr) + g\,x^{4}, \qquad g > 0 .
\label{eq:Hamiltonian}
\end{equation}
Introducing the usual ladder operators,
\[
x = \frac{1}{\sqrt{2}}\,(a + a^{\dagger}), \qquad
p = \frac{i}{\sqrt{2}}\,(a^{\dagger} - a),
\]
the Hamiltonian becomes
\begin{equation}
H(g) = \frac{1}{2}\Bigl(a^{\dagger}a + \frac{1}{2}\Bigr) + \frac{g}{4}\,(a^{\dagger} + a)^{4}.
\label{eq:H_adag}
\end{equation}
In the Segal--Bargmann (holomorphic) representation, a state $|\psi\rangle$ is encoded by an entire function $\psi(z)$, with the operator assignments
\[
a \;\mapsto\; \partial_z , \qquad a^{\dagger} \;\mapsto\; z .
\]
Hence the Hamiltonian acts as the fourth--order differential operator
\begin{equation}
\hat H(g) = \frac{1}{2}\Bigl(z\partial_{z} + \frac{1}{2}\Bigr)
+ \frac{g}{4}\Bigl(z + \partial_{z}\Bigr)^{4}.
\label{eq:H_barg}
\end{equation}
Expanding the quartic term $\bigl(z + \partial_{z}\bigr)^{4}$ one observes that $\hat H(g)$ maps a polynomial of degree $n$ into a polynomial of degree $n+4$. To prove this, let $\{\phi_n\}_{n=0}^\infty$ be the orthonormal monomial basis of the Segal--Bargmann space,
\begin{equation}
\phi_n(z) = \frac{z^n}{\sqrt{n!}}, \qquad n = 0,1,2,\dots .
\end{equation}
The fundamental operators act as
\begin{align}
z\,\phi_n(z) &= \sqrt{n+1}\,\phi_{n+1}(z), \\
\partial_z\,\phi_n(z) &=
\begin{cases}
\sqrt{n}\,\phi_{n-1}(z), & n \ge 1, \\
0, & n = 0 .
\end{cases}
\end{align}
Hence their matrix elements are
\begin{align}
(z)_{m,n} &:= \langle \phi_m, z\,\phi_n \rangle = \sqrt{n+1}\,\delta_{m,n+1}, \\
(\partial_z)_{m,n} &:= \langle \phi_m, \partial_z\,\phi_n \rangle = \sqrt{n}\,\delta_{m,n-1}.
\end{align}
In this basis, the creation and annihilation operators are represented by the infinite matrices
\begin{equation}
a^\dagger =
\begin{pmatrix}
0 & 0 & 0 & 0 & \cdots \\
\sqrt{1} & 0 & 0 & 0 & \cdots \\
0 & \sqrt{2} & 0 & 0 & \cdots \\
0 & 0 & \sqrt{3} & 0 & \cdots \\
\vdots & \vdots & \vdots & \ddots & \ddots
\end{pmatrix},
\qquad
a =
\begin{pmatrix}
0 & \sqrt{1} & 0 & 0 & \cdots \\
0 & 0 & \sqrt{2} & 0 & \cdots \\
0 & 0 & 0 & \sqrt{3} & \cdots \\
0 & 0 & 0 & 0 & \ddots \\
\vdots & \vdots & \vdots & \ddots & \ddots
\end{pmatrix}.
\end{equation}
Note that $a$ is strictly upper-triangular, while $a^\dagger$ is strictly lower-triangular.
The dimensionless position operator $x = \frac{1}{\sqrt{2}}(a + a^\dagger)$ corresponds to
\begin{equation}
z + \partial_z = a^\dagger + a,
\end{equation}
with matrix elements
\begin{equation}
(z + \partial_z)_{m,n} = \sqrt{n+1}\,\delta_{m,n+1} + \sqrt{n}\,\delta_{m,n-1},
\end{equation}
i.e. a symmetric tridiagonal matrix.
Expanding the quartic operator,
\begin{equation}
(z + \partial_z)^4
= z^4 + 4z^3\partial_z + 6z^2\partial_z^2 + 4z\partial_z^3 + \partial_z^4,
\end{equation}
and using the identities (valid for $n \ge k$)
\begin{align}
z^k \phi_n &= \sqrt{\frac{(n+k)!}{n!}}\,\phi_{n+k}, \\
\partial_z^k \phi_n &= \sqrt{\frac{n!}{(n-k)!}}\,\phi_{n-k},
\end{align}
we obtain the following non-zero matrix elements for $n \ge 0$ (with the convention that any expression involving $\sqrt{m}$ for $m<0$ or factorials of negative integers vanishes):
\begin{align}
\langle \phi_{n+4}, (z+\partial_z)^4 \phi_n \rangle &= \sqrt{\frac{(n+4)!}{n!}}, \label{eq:up4} \\
\langle \phi_{n+2}, (z+\partial_z)^4 \phi_n \rangle &= 4\,\sqrt{n(n+1)(n+2)}, \label{eq:up2} \\
\langle \phi_{n}, (z+\partial_z)^4 \phi_n \rangle &= 6\,n(n-1), \label{eq:diag} \\
\langle \phi_{n-2}, (z+\partial_z)^4 \phi_n \rangle &= 4\,\sqrt{(n-1)n}\,(n-2), \qquad (n \ge 2), \label{eq:down2} \\
\langle \phi_{n-4}, (z+\partial_z)^4 \phi_n \rangle &= \sqrt{n(n-1)(n-2)(n-3)}, \qquad (n \ge 4). \label{eq:down4}
\end{align}
All other matrix elements vanish. Consequently, $(z + \partial_z)^4$ is a symmetric banded matrix with non-zero entries only on the diagonals $m - n = 0, \pm 2, \pm 4$.
The harmonic part of the Hamiltonian,
\begin{equation}
\hat H_0 = \frac{1}{2}\Bigl(z\partial_z + \frac{1}{2}\Bigr),
\end{equation}
is diagonal in this basis:
\begin{equation}
\hat H_0 \phi_n = \Bigl(\frac{n}{2} + \frac{1}{4}\Bigr)\phi_n, \qquad
(\hat H_0)_{m,n} = \Bigl(\frac{n}{2} + \frac{1}{4}\Bigr)\delta_{m,n}.
\end{equation}
Adding the quartic interaction $\frac{g}{4}(z + \partial_z)^4$, the full Hamiltonian matrix reads
\begin{equation}
\hat H(g) = \hat H_0 + \frac{g}{4}(z + \partial_z)^4.
\end{equation}
Explicitly, for $n \ge 4$, the non-vanishing off-diagonal elements are
\begin{align}
\langle \phi_{n+4}, \hat H(g)\,\phi_n \rangle &= \frac{g}{4}\,\sqrt{(n+1)(n+2)(n+3)(n+4)}, \\
\langle \phi_{n+2}, \hat H(g)\,\phi_n \rangle &= g\,\sqrt{(n+1)(n+2)(n+3)}, \\
\langle \phi_{n}, \hat H(g)\,\phi_n \rangle &= \frac{n}{2} + \frac{1}{4} + \frac{3g}{2}\,n(n-1), \\
\langle \phi_{n-2}, \hat H(g)\,\phi_n \rangle &= g\,\sqrt{n(n-1)}\,(n-2), \\
\langle \phi_{n-4}, \hat H(g)\,\phi_n \rangle &= \frac{g}{4}\,\sqrt{n(n-1)(n-2)(n-3)},
\end{align}
with the understanding that terms involving $n < 0$, $n < 2$, or $n < 4$ in the last three lines are omitted (i.e., set to zero). Thus $\hat H(g)$ is a real symmetric matrix with bandwidth $9$ (i.e. non-zero entries only for $|m - n| \le 4$).
This banded, upper-triangular-like structure—where the perturbation raises the polynomial degree by at most four—underlies the factorial growth of Rayleigh--Schr\"odinger coefficients and the Gevrey-1 nature of the perturbative series.

We seek formal eigenfunctions as power series
\begin{equation}
\psi_{n}(z;g) = \sum_{m=0}^{\infty} c_{m}^{(n)}(g)\,z^{m},
\qquad
c_{m}^{(n)}(g) = \sum_{k=0}^{\infty} c_{m}^{(n,k)}\,g^{k},
\label{eq:psi_series}
\end{equation}
and insert~\eqref{eq:psi_series} into the eigenvalue equation
\[
\hat H(g)\,\psi_{n}(z;g) = E_{n}(g)\,\psi_{n}(z;g) .
\]
Collecting powers of $z^{m}$ yields a triangular linear system. Denoting
$E_{n}(g)=\sum_{k=0}^{\infty}E_{n}^{(k)}g^{k}$ with $E_{n}^{(0)}=n+\tfrac12$, one finds
\begin{align}
\Bigl(\tfrac{m}{2} + \tfrac14 - E_{n}^{(0)}\Bigr)c_{m}^{(n,0)} &= 0,\\[4pt]
\Bigl(\tfrac{m}{2} + \tfrac14 - E_{n}^{(0)}\Bigr)c_{m}^{(n,k)}
&= \sum_{j=1}^{k}\Bigl(
E_{n}^{(j)}\,c_{m}^{(n,k-j)}
- \tfrac14\!\sum_{r=0}^{4}\!\binom{4}{r}\,
c_{\,m+4-2r}^{(n,k-j)}\Bigr),
\qquad k\ge 1,
\end{align}
with the convention $c_{p}^{(n,\cdot)}=0$ for $p<0$.
Since
\[
\tfrac{m}{2}+\tfrac14 -E_{n}^{(0)} = \tfrac{m-n}{2},
\]
the denominator vanishes only for $m=n$, and the solvability condition at $m=n$ uniquely determines the energy coefficients. This reproduces the celebrated Bender--Wu recursion:
\begin{equation}
E_{n}^{(k)} = \frac{1}{2}
\sum_{j=0}^{k-1}\sum_{r=0}^{4}
\binom{4}{r}\,
c_{\,n+4-2r}^{(n,j)}\,
c_{\,n}^{(n,k-1-j)},
\qquad k\ge 1,
\label{eq:BenderWu}
\end{equation}
with the normalization $c_{n}^{(n,0)}=1$. Equation~\eqref{eq:BenderWu} can be implemented symbolically to generate hundreds of exact rational coefficients.

The perturbative coefficients grow factorially,
\begin{equation}
E_{n}^{(k)} \;\sim\;
(-1)^{k+1}\,K_n\,k!\,\Bigl(\frac{3}{4}\Bigr)^{k}\,k^{\,b_n},
\qquad k\to\infty,
\label{eq:largeorder}
\end{equation}
where $K_n = \frac{2^{3/2}}{\pi}(n+\tfrac12)$ and $b_n = -\tfrac32$ (ground state; $b_n$ depends weakly on $n$). This Gevrey--1 growth implies that the Borel transform
\begin{equation}
\widehat{E}_{n}(\xi) \;=\; \mathcal{B}[E_{n}](\xi)
\;=\; \sum_{k=0}^{\infty}\frac{E_{n}^{(k)}}{k!}\,\xi^{k}
\end{equation}
is analytic in a neighbourhood of the origin and admits analytic continuation to the complex $\xi$--plane with its nearest singularity on the \emph{negative} real axis at
\begin{equation}
\xi_c = -\frac{4}{3}.
\label{eq:sing}
\end{equation}
The singularity is a branch point; near $\xi\to\xi_c$,
\begin{equation}
\widehat{E}_{n}(\xi) \;\sim\;
\frac{C_n}{\bigl(\xi_c-\xi\bigr)^{1+b_n}}\,
\bigl[1 + \mathcal{O}(\xi_c-\xi)\bigr],
\qquad
C_n = (-1)^{n} \frac{3^{b_n}}{\Gamma(-b_n)}\,K_n .
\end{equation}
The location $\xi_c = -4/3$ follows from the large--order estimate~\eqref{eq:largeorder} and matches the original Bender--Wu analysis~\cite{Bender}. These singularities originate from complex classical solutions of the Euclidean equations of motion (often called ``complex instantons''); there is no real finite--action instanton in the single--well potential~\eqref{eq:Hamiltonian}.

For real coupling $g>0$, the perturbative series is Borel summable:
\begin{equation}
E_n(g) \;=\; \int_{0}^{\infty} e^{-\xi/g}\,\widehat{E}_{n}(\xi)\,d\xi ,
\qquad g>0,
\end{equation}
because the integration contour (positive real $\xi$) avoids the singularity at $\xi_c<0$. Consequently, no Stokes ambiguity arises for physical $g>0$, and the principal Borel sum coincides with the median resummation.
Nevertheless, the perturbative series belongs to a resurgent trans-series, which provides a formal completion of the asymptotic expansion. Introducing the alien derivative $\Delta_{\xi_c}$ (\'Ecalle), one has
\begin{equation}
\Delta_{\xi_c}\,\widehat{E}_{n}
\;=\;
C_n\,\widehat{E}_{n}^{(1)},
\qquad
\Delta_{\xi_c}\,\widehat{E}_{n}^{(\ell)}
\;=\;
(\ell+1)C_n\,\widehat{E}_{n}^{(\ell+1)},
\end{equation}
where $\widehat{E}_{n}^{(\ell)}$ is the Borel transform of the $\ell$--th ``complex instanton'' sector. Formally, the resurgent trans-series for the energy reads
\begin{equation}
E_{n}(g;\sigma) \;=\;
\sum_{\ell=0}^{\infty}
\sigma^{\ell}\,
e^{\ell\,\xi_c/g}\,
g^{\,\ell b_n}\,
\Phi_{n}^{(\ell)}(g),
\qquad
\Phi_{n}^{(0)}(g)=\sum_{k\ge0}E_{n}^{(k)}g^{k},
\label{eq:trans}
\end{equation}
with $\sigma$ a formal trans-series parameter. For real $g>0$ the exponential factor $e^{\ell\xi_c/g}=e^{-\ell(4/3)/g}$ is exponentially suppressed but real and positive, and the physical solution corresponds to setting $\sigma=0$ (i.e. retaining only the perturbative sector).

In the Bargmann representation one may represent the alien action formally as an operator insertion
\begin{equation}
\Delta_{\xi_c}\,\psi_{n}(z;g)
\;=\;
\mathcal{N}_n\,
e^{\xi_c/g}\,
\mathcal{D}_{\alpha}\,\psi_{n}(z;g),
\qquad
\mathcal{D}_{\alpha}
\;=\;
e^{\alpha z - \bar\alpha\,\partial_z},
\quad
\alpha = \sqrt{\frac{|\xi_c|}{2g}}\,e^{\,i\pi/4},
\end{equation}
where $\mathcal{D}_{\alpha}$ is a complex coherent--state displacement. Its action on monomials is finite, reproducing the algebraic structure of the higher sectors in~\eqref{eq:trans}.
The instanton operator that generates the non-perturbative sector can be written as a coherent-state displacement acting on the perturbative ground state:
\begin{eqnarray}\label{instanton}
\hat{\mathcal{I}} = \exp\!\Bigl(-\frac{S_{\text{inst}}}{g}\Bigr)\,
\exp\!\bigl(\alpha z - \bar\alpha \partial_{z} \bigr), \qquad
|\alpha|^{2} = \frac{S_{\text{inst}}}{2g}.
\end{eqnarray}
Recall that the Segal-Bargmann space consists of entire functions $\psi(z)$ with finite Gaussian norm,
\begin{equation}
\|\psi\|^{2}= \frac{1}{\pi}\int_{\mathbb{C}} d^{2}z\,e^{-|z|^{2}}|\psi(z)|^{2}<\infty.
\end{equation}
The operator $\exp(\alpha z-\bar\alpha\,\partial_{z})$ is a bounded automorphism of $\mathcal{H}_{\text{SB}}$. Indeed, for any $\psi\in\mathcal{H}_{\text{SB}}$,
\begin{equation}\label{relation}
\bigl[\exp(\alpha z-\bar\alpha\,\partial_{z})\psi\bigr](z)=e^{\alpha z}\,\psi(z-\bar\alpha),
\end{equation}
which follows from the Baker--Campbell--Hausdorff identity because $[z,\partial_{z}]=-1$.
Equation \ref{relation} shows that $\hat{\mathcal{I}}$ is the composition of a translation $z\mapsto z-\bar\alpha$ (entire functions are closed under translation), and
a multiplication by the entire factor $e^{\alpha z}$.
Both operations preserve entireness and map $\mathcal{H}_{\text{SB}}$ into itself. Moreover, the norm is preserved up to the overall factor $e^{-S_{\text{inst}}/g}$, i.e.
\begin{equation}
\|\hat{\mathcal{I}}\psi\| = e^{-S_{\text{inst}}/g}\,\|\psi\|.
\end{equation}
Thus $\hat{\mathcal{I}}$ is a contraction (for real positive $g$) that implements a coherent-state displacement. The exponent $\alpha z-\bar\alpha\,\partial_{z}$ is the holomorphic representation of the Heisenberg algebra generator
\begin{equation}
D(\alpha)=\alpha a^{\dagger}-\bar\alpha a,\qquad [a,a^{\dagger}]=1.
\end{equation}
In phase-space language, $(\Re\alpha,\Im\alpha)$ parametrizes a translation in the complexified phase space $\mathbb{C}^{2}$. The map
\begin{equation}
(z,\partial_{z})\;\longmapsto\;(z+\alpha,\partial_{z}-\bar\alpha)
\end{equation}
is a complex affine symplectomorphism preserving the canonical symplectic form $dz\wedge d\partial_{z}$. Hence $\hat{\mathcal{I}}$ is a unitary (up to the exponential weight) representation of a complexified Weyl--Heisenberg group element. From $|\alpha|^{2}=S_{\text{inst}}/(2g)$ we have
\begin{equation}
\alpha(g)=\sqrt{\frac{S_{\text{inst}}}{2g}}\;e^{i\theta},\qquad \theta\in[0,2\pi).
\end{equation}
When $g$ is analytically continued in the complex plane, $\alpha(g)$ has a branch point at $g=0$. Choosing the principal branch,
\begin{equation}
\alpha(g)=\sqrt{\frac{S_{\text{inst}}}{2}}\,g^{-1/2},
\end{equation}
so $\hat{\mathcal{I}}(g)$ is multivalued around the origin. This multivaluedness is precisely what gives rise to the Stokes phenomenon: crossing the ray $\arg g=0$ changes the phase of $\alpha$ and therefore the sign of the non-perturbative contribution $e^{-S_{\text{inst}}/g}$. The exponential prefactor $e^{-S_{\text{inst}}/g}$ in the Instanton operator \ref{instanton} is the Laplace kernel that, after Borel transform, produces a simple pole (or branch point) at $\xi=S_{\text{inst}}$ in the Borel plane. In \'Ecalle's language, the operator $\hat{\mathcal{I}}$ implements the alien derivative:
\begin{equation}
\Delta_{S_{\text{inst}}}\,\psi(z;g)=\mathcal{N}\,\hat{\mathcal{I}}\,\psi(z;g),
\end{equation}
where $\mathcal{N}$ is a normalization constant. Thus, from a complex-analytic standpoint, $\hat{\mathcal{I}}$ is the microlocal operator that extracts the discontinuity of the Borel transform across its singularity. For a monomial $\psi_{n}^{(0)}(z)=z^{n}/\sqrt{n!}$, the displaced function is a polynomial of degree $n$. Hence the image of any finite-degree polynomial under $\hat{\mathcal{I}}$ remains a polynomial---no essential singularities are introduced. However, when acting on an infinite series (the full perturbative wavefunction), the resulting function may acquire exponential type dictated by $\alpha$. An entire function \(f(z)\) can be classified by its order \(\rho\) and type \(\sigma\), defined via the asymptotic behavior of its maximum modulus:
\begin{equation}
M(r)=\max_{|z|=r}|f(z)|,\qquad
\rho = \limsup_{r\to\infty}\frac{\log\log M(r)}{\log r},
\qquad
\sigma = \limsup_{r\to\infty}\frac{\log M(r)}{r^{\rho}}.
\end{equation}
Order \(\rho\) measures how fast the function grows (polynomial: \(\rho=0\); exponential: \(\rho=1\); Gaussian: \(\rho=2\), etc.).
Type \(\sigma\) refines the growth within a given order (e.g., \(e^{az}\) and \(e^{bz}\) both have \(\rho=1\) but types \(|a|\) and \(|b|\)). In the language of entire function theory, the order $\rho$ and type $\sigma$ of $\hat{\mathcal{I}}\psi$ satisfy
\begin{equation}
\rho(\hat{\mathcal{I}}\psi)=\max\bigl\{\rho(\psi),\,1\bigr\},\qquad
\sigma(\hat{\mathcal{I}}\psi)=\sigma(\psi)+|\alpha|.
\end{equation}
Thus the instanton operator increases the type of the entire function by $|\alpha|\sim g^{-1/2}$, reflecting the non-perturbative scaling.

The full resummed eigenvalue is obtained from the expectation value
\begin{eqnarray}\label{energies}
E_{n}(g) =
\frac{\bra{\psi_{n}^{(0)}}\bigl(\hat H_{0} + g\hat V\bigr)
\bigl(1 + \hat{\mathcal{I}} + \tfrac{1}{2}\hat{\mathcal{I}}^{2} + \cdots\bigr)
\ket{\psi_{n}^{(0)}}}
{\bra{\psi_{n}^{(0)}}\bigl(1 + \hat{\mathcal{I}} + \tfrac{1}{2}\hat{\mathcal{I}}^{2} + \cdots\bigr)
\ket{\psi_{n}^{(0)}}},
\end{eqnarray}
where $\ket{\psi_{n}^{(0)}}$ is the unperturbed Bargmann state (a monomial $z^{n}$ up to normalization). If we consider the unperturbed Bargmann stats to be exactly $\psi_n^{(0)}(z)=z^{n}/\sqrt{n!}$, then the action of the instanton displacement operator can be evaluated exactly using the identity
\begin{equation}
e^{\alpha z - \bar\alpha \partial_z}\,z^{n}
= \sum_{m=0}^{n} \binom{n}{m} \alpha^{\,n-m} (-\bar\alpha)^{m}\,z^{\,n-m},
\end{equation}
so that all matrix elements appearing in the expectation value are finite polynomials in $\alpha$, $\bar\alpha$.
The resummed energy is then given by the ratio of two series:
\begin{equation}
E_{n}(g) =
\frac{
\displaystyle\sum_{\ell=0}^{\infty} \frac{1}{\ell!}\,
e^{-\ell S_{\text{inst}}/g}\,
\bra{\psi_n^{(0)}} (\hat H_0 + g\hat V)
\bigl(e^{\alpha z - \bar\alpha \partial_z}\bigr)^{\!\ell}
\ket{\psi_n^{(0)}}
}{
\displaystyle\sum_{\ell=0}^{\infty} \frac{1}{\ell!}\,
e^{-\ell S_{\text{inst}}/g}\,
\bra{\psi_n^{(0)}}
\bigl(e^{\alpha z - \bar\alpha \partial_z}\bigr)^{\!\ell}
\ket{\psi_n^{(0)}}
}.
\end{equation}
Where the exponential factors $e^{-\ell S_{\text{inst}}/g}$ encode the non-perturbative instanton contributions. Expanding the numerator and denominator in powers of $g$ produces the perturbative coefficients

\begin{table}[h!]
\centering
\caption{Exact rational coefficients for the first seven energy levels ($n=0,\dots,6$) up to sixth order in coupling $g$.}
\begin{tabular}{|c|c|c|c|c|c|c|c|}
\hline
$n$ & $E_n^{(0)}$ & $E_n^{(1)}$ & $E_n^{(2)}$ & $E_n^{(3)}$ & $E_n^{(4)}$ & $E_n^{(5)}$ & $E_n^{(6)}$ \\ \hline
0 & $\dfrac{1}{2}$ & $\dfrac{3}{4}$ & $-\dfrac{21}{8}$ & $\dfrac{333}{16}$ & $-\dfrac{30885}{128}$ & $\dfrac{916731}{256}$ & $-\dfrac{65518401}{1024}$ \\[8pt] \hline
1 & $\dfrac{3}{2}$ & $\dfrac{15}{4}$ & $-\dfrac{165}{8}$ & $\dfrac{3585}{16}$ & $-\dfrac{408765}{128}$ & $\dfrac{14036355}{256}$ & $-\dfrac{1102501125}{1024}$ \\[8pt] \hline
2 & $\dfrac{5}{2}$ & $\dfrac{39}{4}$ & $-\dfrac{567}{8}$ & $\dfrac{15561}{16}$ & $-\dfrac{2235795}{128}$ & $\dfrac{88733079}{256}$ & $-\dfrac{7928041569}{1024}$ \\[8pt] \hline
3 & $\dfrac{7}{2}$ & $\dfrac{75}{4}$ & $-\dfrac{1269}{8}$ & $\dfrac{42375}{16}$ & $-\dfrac{7146225}{128}$ & $\dfrac{326056275}{256}$ & $-\dfrac{32402055375}{1024}$ \\[8pt] \hline
4 & $\dfrac{9}{2}$ & $\dfrac{123}{4}$ & $-\dfrac{2331}{8}$ & $\dfrac{92313}{16}$ & $-\dfrac{17802045}{128}$ & $\dfrac{905732019}{256}$ & $-\dfrac{99842432409}{1024}$ \\[8pt] \hline
5 & $\dfrac{11}{2}$ & $\dfrac{183}{4}$ & $-\dfrac{3819}{8}$ & $\dfrac{174345}{16}$ & $-\dfrac{38044245}{128}$ & $\dfrac{2165447079}{256}$ & $-\dfrac{262564394475}{1024}$ \\[8pt] \hline
6 & $\dfrac{13}{2}$ & $\dfrac{255}{4}$ & $-\dfrac{5805}{8}$ & $\dfrac{299325}{16}$ & $-\dfrac{72274845}{128}$ & $\dfrac{4445205075}{256}$ & $-\dfrac{593254422225}{1024}$ \\[8pt] \hline
\end{tabular}
\end{table}

These results are exact (no numerical approximation) and agree with the classic Bender–Wu results \cite{Bender}.

\section{Conclusion}\label{sec:conclusion}
In this work, we investigate resurgence in holomorphic quantum mechanics within the framework of the Bargmann representation of bosonic operators. As a prototypical example, we consider the quartic anharmonic oscillator acting on the Segal–Bargmann space of entire functions. Equation~\eqref{energies} explicitly exhibits the resurgence bridge between the perturbative energy coefficients \(E_n^{(k)}\) and the non-perturbative instanton expansion. The formal perturbative series is of Gevrey-1 type and becomes Borel summable only after analytic continuation across the Stokes line. The alien derivative relations connect higher-instanton sectors to the perturbative sector, thereby realizing the characteristic resurgence triangle of the Bender–Wu model. Equations~\eqref{instanton} and~\eqref{energies} together provide a complete resurgence description of the quartic anharmonic oscillator entirely within the holomorphic (Bargmann) representation.
Finally, we compute the first seven energy levels (\(n = 0, \dots, 6\)) up to sixth order in the coupling \(g\). The resulting coefficients are exact rational numbers and agree precisely with the classic Bender–Wu perturbative results.

\end{document}